\documentclass[journal=ancac3,manuscript=article]{achemso}
\setkeys{acs}{keywords = true}

\usepackage{amssymb,amsfonts,amsmath}
\usepackage{graphicx}
\usepackage{indentfirst}
\usepackage{setspace}
\title{Voltage Controlled Interfacial Layering in an Ionic Liquid on SrTiO$_3$} 

\author{Trevor Petach}
\affiliation{Department of Physics, Stanford University, Palo Alto, CA 94305, USA}
\author{Apurva Mehta}
\affiliation{SLAC National Accelerator Laboratory, Menlo Park, CA 94025, USA}
\author{Ronald Marks}
\affiliation{SLAC National Accelerator Laboratory, Menlo Park, CA 94025, USA}
\author{Bart Johnson}
\affiliation{SLAC National Accelerator Laboratory, Menlo Park, CA 94025, USA}
\author{Michael Toney}
\affiliation{SLAC National Accelerator Laboratory, Menlo Park, CA 94025, USA}
\author{David Goldhaber-Gordon}
\email{goldhaber-gordon@stanford.edu}
\phone{+1 (650) 724-3681}
\affiliation{Department of Physics, Stanford University, Palo Alto, CA 94305, USA}


\begin{document}






\begin{abstract}
One prominent structural feature of ionic liquids near surfaces is formation of alternating layers of anions and cations. However, how this layering responds to applied potential is poorly understood. We focus on the structure of 1-butyl-1-methylpyrrolidinium tris(pentafluoroethyl) trifluorophosphate (BMPY-FAP) near the surface of a strontium titanate (SrTiO$_3$) electric double-layer transistor. Using x-ray reflectivity, we show that at positive bias, the individual layers in the ionic liquid double layer thicken and the layering persists further away from the interface. We model the reflectivity using a modified distorted crystal model with alternating cation and anion layers, which allows us to extract the charge density and the potential near the surface. We find that the charge density is strongly oscillatory with and without applied potential, and that with applied gate bias of 4.5 V the first two layers become significantly more cation rich than at zero bias, accumulating about $2.5 \times 10^{13}$ cm$^{-2}$ excess charge density.
\end{abstract}

\section*{}
Ionic liquids are widely used in electrochemical systems, including electric double-layer transistors (EDLTs) \cite{Yuan2009,Petach2014}, electrodeposition at high potentials \cite{Endres2008,Armand2009,Tuodziecki2014}, and supercapacitors \cite{Kim2005a}. The arrangement of ions near the electrodes plays a key role in these systems. For example, separating the ionic liquid slightly from the channel surface using an insulating spacer increases mobility by an order of magnitude in electrolyte gated SrTiO$_3$ \cite{Gallagher2015}, and the grain size and shape in films electrodeposited from ionic liquid electrolytes depends on which ionic liquid is used \cite{ZeinElAbedin2006,Borisenko2006}.

The arrangement of ions in the ionic liquid double-layer is governed by a steric effect from the finite size of the ions and a coulomb effect from the charge of the ions. In aqueous solutions, where the constituent water molecules have no net charge, alternating light and dense layers have been observed on Ag (111) \cite{Toney1994}, SrTiO$_3$ (001) \cite{Plaza2015}, and mineral surfaces \cite{Fenter2005,Eng2000}. This layering is a universal feature of hard spheres near a hard wall, in which some spheres tend to form a layer on the wall, forcing the spheres just away from the wall to also form layers \cite{Watanabe2011a}. In ionic liquids, the strong coulomb interaction between the anions and cations also plays a role in the interfacial structure. This interaction should give rise to additional effects such as alternating cation layers and anion layers \cite{Kornyshev2007}.

Indeed, formation of alternating out-of-plane layers of anions and cations has been observed using x-ray reflectivity \cite{Mezger2008}. The strength and extent of the layering depends on the substrate - liquid interaction and can be modeled using modified molecular dynamics simulations \cite{Zhou2012a}. However, how this layering responds to applied potential is less well studied. Atomic force-distance profiles show distinct plateaus in force as a function of distance from the electrode across a broad range of potentials \cite{Hayes2011}, suggesting some sort of layering, but the presence of the tip most likely perturbs the double layer structure, and it is hard to quantitatively obtain ion concentrations in each layer from force profiles. X-ray reflectivity has been used to study ionic liquids on gold surfaces at several potentials \cite{Yamamoto2012}, but that study was unable to distinguish between layering confined to 1 or 2 layers near the electrode or extended further into the liquid. Lateral ordering of ions has also been observed \cite{Elbourne2015}, but in this work we focus on out-of-plane ordering.

We focus on the structure of the ionic liquid double layer in a SrTiO$_3$ (001) electric double layer transistor at positive gate potentials. SrTiO$_3$ transitions from an insulating state at low gate bias to a metallic state at positive bias \cite{Lee2011f}. Using x-ray reflectivity, we show that the ionic liquid near the interface forms alternating cation-rich and anion-rich layers at all gate potentials studied. Coincident with the electronic change in the SrTiO$_3$ at positive bias, the spatial period of the layering becomes longer and the layering persists further away from the interface. Using the reflectivity to constrain a model for electron density near the interface, we construct a picture of three aspects of the double-layer in this system: the charge density, potential, and interfacial capacitance.

These results are a quantitative description of the structure of the interface under bias, a critical step to develop effectual theories of the ionic liquid double layer.

\begin{figure*}[hp!]
\includegraphics[width=\columnwidth]{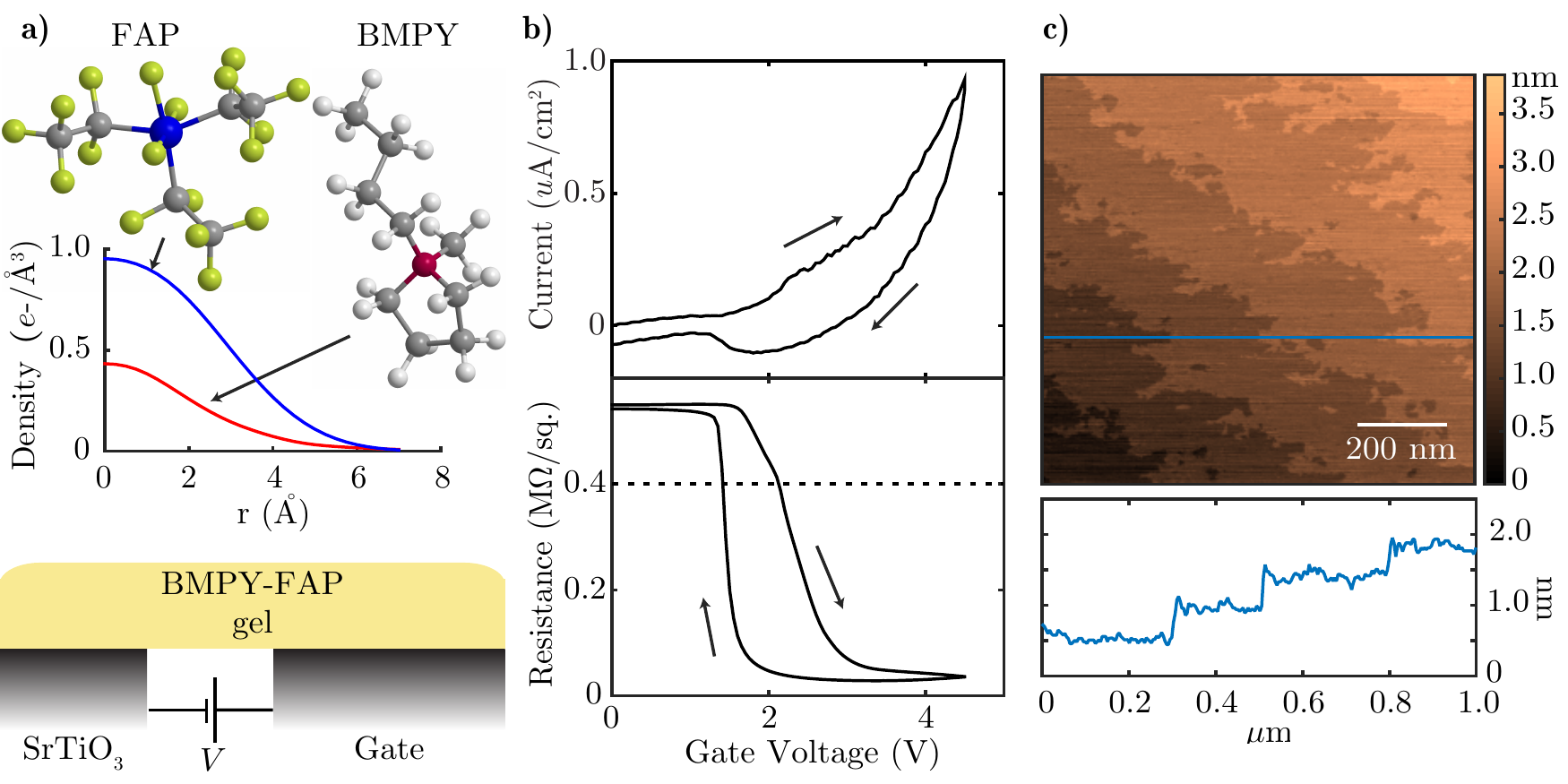}
\caption{The electric double layer transistor. (a) The ionic liquid used in this study. BMPY is 1-butyl-1-methylpyrrolidinium, and FAP is tris(pentafluoroethyl)trifluorophosphate. The total radial electron density (about the center of mass) is plotted for BMPY (red) and FAP (blue). The liquid was spread in a thin layer across a gate electrode and the SrTiO$_3$. (b) Current density (top) and sheet resistance (bottom) during cyclic voltammetry, sweep rate 50 mV/s. Current density is normalized to the surface area of the SrTiO$_3$ working electrode (1 cm$^2$). The gold counter electrode is also 1 cm$^2$. The dashed line indicates the approximate resistance of the ionic liquid film, above which the measurement becomes unreliable. (c) Atomic force micrograph of the SrTiO$_3$ channel after surface preparation described in the text.}
\label{fig:overview}
\end{figure*}

\subsection*{Results and Discussion}

\begin{figure}[hp!]
\includegraphics[width=3.375in]{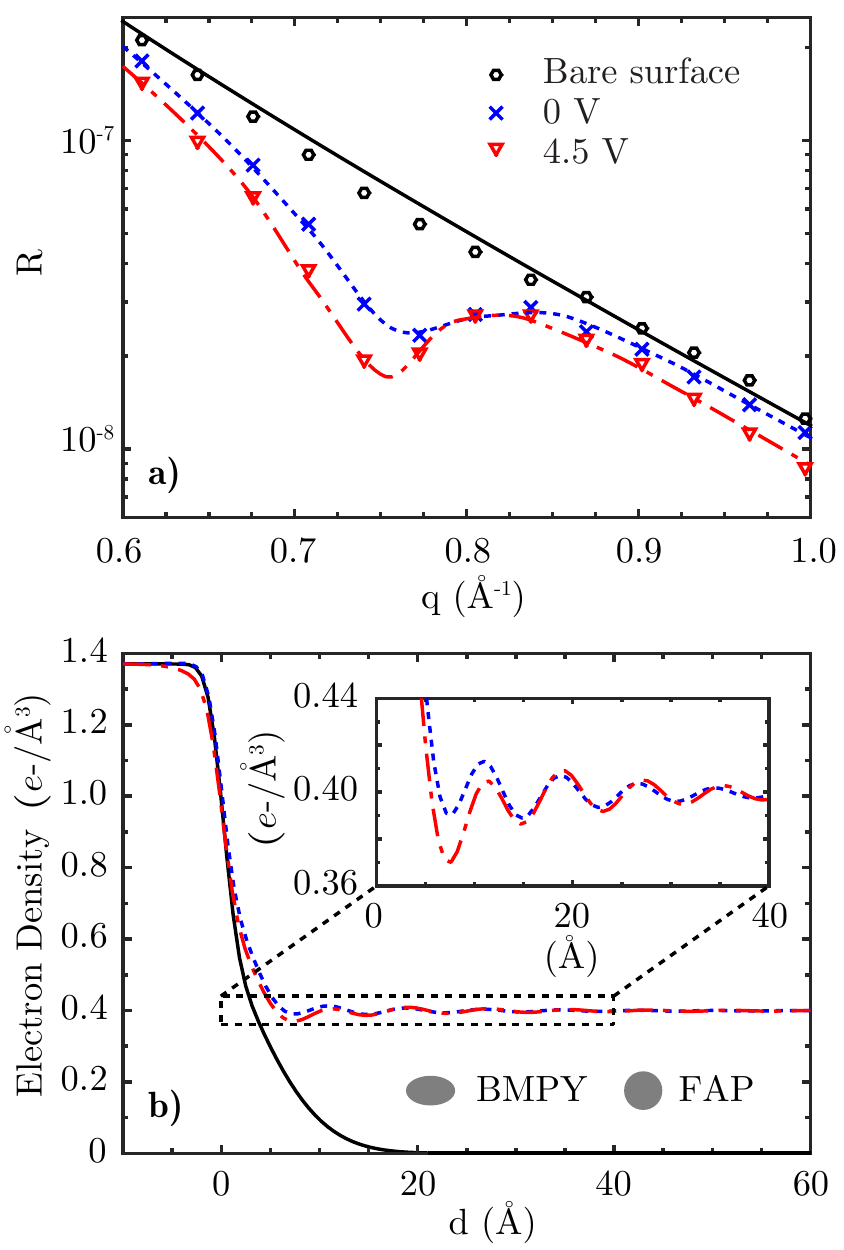}
\caption{Formation of layered structure in BMPY-FAP at the SrTiO$_3$ surface. (a) Reflectivity. Symbols are data and lines are best fit models. (b) The best fit electron density profile. The approximate size of the ions is shown for reference.}
\label{fig:XRR}
\end{figure}

Figure \ref{fig:XRR} shows the reflected intensity as a function of scattering vector. A background, which was obtained from pixels adjacent to the reflectivity peak on the area detector, has been subtracted from the data. Without the ionic liquid, the reflected intensity is almost featureless. With BMPY-FAP, a dip near 0.77 \AA$^{-1}$ develops. With positive applied bias, this dip deepens and shifts toward smaller scattering vectors.

The isolated dip (see Supporting Information for a figure showing reflectivity over a larger $q$ range) suggests destructive interference from multiple layers of similar thickness. The deepening of the dip at positive bias suggests that the electron density contrast between the layers increases or that the number of layers increases, and the shifting of the dip to lower $q$ suggests that the individual layers become thicker. These observations guided development of a model for the electron density near the interface.

The data were fit using a genetic algorithm that minimized the quantity $\sum | \log(R_{model}) - \log(R_{data}) | ^ 2$ by optimizing the parameters in a model for electron density. Parratt's method \cite{Parratt1954} was used to calculate the reflected intensity from the electron density. The model for the bare surface consisted of a substrate, whose roughness was allowed to vary, and a surface layer, whose thickness, density, and roughness were allowed to vary. The surface layer, which is light and thin, accounts for the small amount of organic material present on all surfaces exposed to ambient conditions. The model for samples covered with ionic liquid consisted of three parts: first, a substrate, whose roughness was allowed to vary; second, an interfacial layer to describe the transition from substrate to liquid, whose thickness, density, and roughness were allowed to vary; third, a series of alternating light and dense layers in the liquid whose density difference decays to zero far from the interface, parameterized by the distorted crystal formula \cite{Magnussen1995}. The distorted crystal formula has seven parameters which were varied during the fit. The model also included corrections for beam shape, sample size, a slight twotheta offset, and an overall scale factor, which were not varied during the fitting procedure (see Supporting Information for a complete explanation of the models and a list of best fit parameter values).

To investigate whether all of the fitting parameters were necessary to achieve a good fit, two more simplistic models were investigated for the samples covered with ionic liquid: first, a model without the interfacial layer; second, a model without the alternating distorted crystal layers but with 2 interfacial layers. As shown in the Supporting Information, fits using these models are significantly worse than the fit using the model described above, providing confidence in our analysis.

The best fit electron density is shown in Figure \ref{fig:XRR}(b). The BMPY cation has low electron density and the FAP cation has high electron density (see Figure \ref{fig:overview}(a)), so the dips in electron density correspond to cation-rich layers and the peaks correspond to anion-rich layers. As expected from alternating layers with primarily cations in one and primarily anions in the next, the spacing between peaks is approximately equal to the sum of the diameters of the BMPY and FAP ions. The best fit peak-to-peak spacing is 7.7 \AA \hspace{1ex}at zero bias to 8.1 \AA \hspace{1ex}at 4.5 V positive bias. The increased layer spacing may be due to rotation of the cations in the cation layers so that the butyl group ``tails'' are perpendicular to the surface, which would increase the separation between anion layers. The electron poor region near the interface at positive bias is due to accumulation of cations on the SrTiO$_3$ surface.

\begin{figure}[hp!]
\includegraphics[width=3.375in]{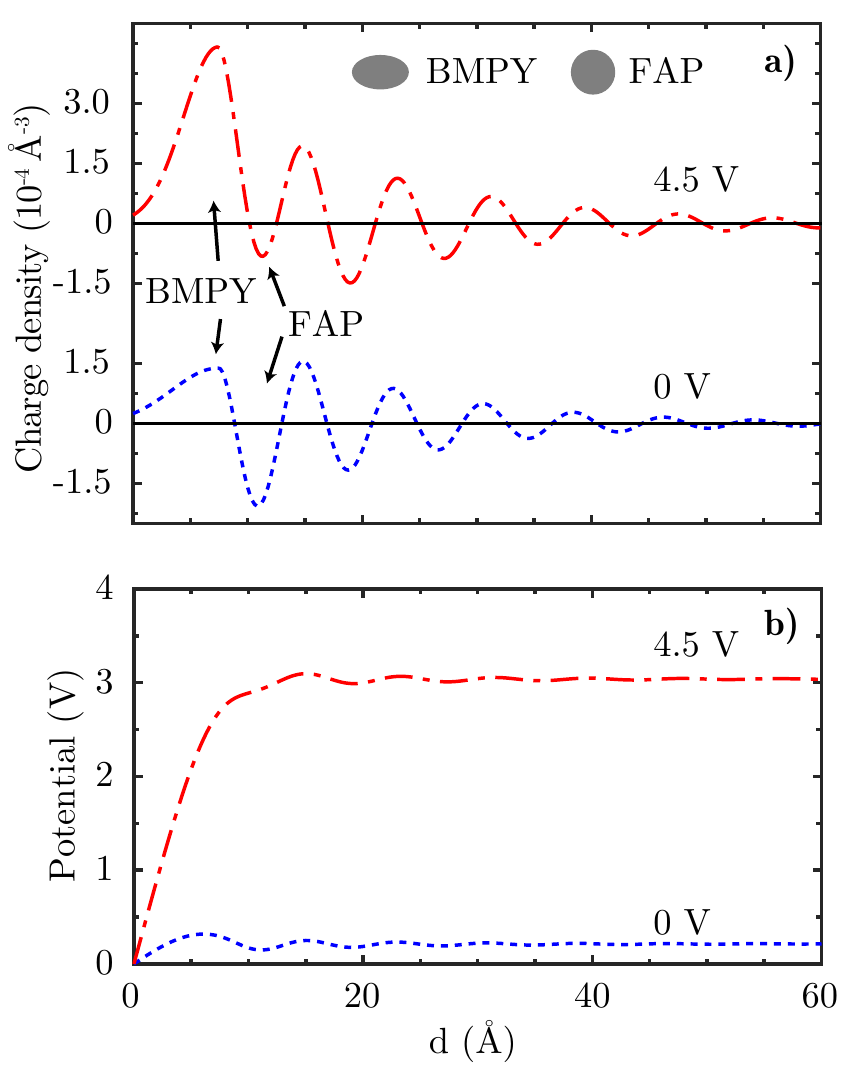}
\caption{Evolution of the interface with applied bias. (a) Net charge density and (b) potential as a function of distance from the interface, calculated from the electron density profile in Figure \ref{fig:XRR}(b). To preserve charge neutrality, we presume a layer of carriers in the SrTiO$_3$ at $d = 0$. The approximate size of the ions is shown for reference.}
\label{fig:Potential}
\end{figure}

By making several simplifying assumptions, the net charge density, and thus the potential, can be calculated from the best fit electron density. Since the average molecular diameter of the cation and anion differ by only 25\%, and they are both approximately spherical (see Figure \ref{fig:overview}(a)), we treat them as spheres of equal diameter in order to calculate the charge density from the electron density (ignoring the substrate). Further presuming that the excess charge in the ionic liquid is compensated by a thin sheet of carriers in the SrTiO$_3$ (located at d = 0 \AA \hspace{1ex}and of equal magnitude and opposite sign as the total charge in the ionic liquid), and that the ions themselves have no dielectric susceptibility, the potential can be calculated from Poisson's equation.

The charge density and potential are shown in Figure \ref{fig:Potential}. The total charge density in the ionic liquid layers is $0.5 \times 10^{13}$ cm$^{-2}$ at 0 V and $2.5 \times 10^{13}$ cm$^{-2}$ at 4.5 V. At both potentials the charge density is strongly oscillatory with distance, a phenomenon known as overscreening in which the interface is characterized by alternating layers of positive and negative charge \cite{Bazant2011}. Overscreening arises from an interplay between ion density, ion size, ion shape, and the Coulomb interaction between the ions \cite{Fedorov2014}. A phenomenon known as crowding occurs when several consecutive layers have an excess of the same ion. 

Though we do not observe crowding, the charge density profile at 4.5 V cannot be explained solely by overscreening. In pure overscreening, the layer closest to the interface should have the highest density of cations and the second layer should have the highest density of anions, with densities in subsequent layers decaying towards the average density in the bulk liquid. As shown in Figure \ref{fig:Potential}(a), the first peak at d = 7 \AA \hspace{1ex}is the most positive, so the first layer indeed has the highest density of cations. However, the trough at d = 11 \AA \hspace{1ex}is relatively shallow compared to troughs farther from the interface, showing that the there are more cations in this layer than would be expected from overscreening alone.

Interestingly, though the charge density is strongly oscillatory at 4.5 V, the potential is not. Most of the potential change across the interface occurs between the surface and the first layer of ions. Also, of the 4.5 V applied between the gate and the channel, only 3 V drop between the bulk liquid and the SrTiO$_3$ surface, indicating that the screening layer of carriers in the SrTiO$_3$ may have finite thickness and that there may be some potential drop between the counter electrode and the bulk ionic liquid. By dividing the total charge in the ionic liquid layers by the potential drop from the bulk liquid to the SrTiO$_3$ surface at 4.5 V gate bias, we calculate the capacitance to be 1.5 $\mu$F/cm$^2$.

Our results further the understanding of fundamental characteristics of ionic liquid double layers. Realistic physical models for ionic liquid double layers are still being developed. Recent progress has been made using molecular dynamics simulations with charged hard spheres of different sizes \cite{Ivanistsev2015}. At low potentials, these simulations predict oscillatory overscreening, and at high potentials, they predict crowding. Our results agree qualitatively with this simulation: we observe overscreening at zero bias, and we observe an increase in cation concentration in two consecutive layers at 4.5 V bias. We also observe additional features, such as layer thickening, that are not captured in these simulations, suggesting that the details of the ion shapes and the region of the ion on which the charge is localized play an important role in the formation of ionic liquid interfacial layers.

\subsection*{Conclusions}

We used x-ray reflectivity to determine the charge and potential profiles in BMPY-FAP as a function of distance from an interface with SrTiO$_3$. We estimated the interfacial capacitance (1.5 $\mu$F/cm$^2$) from a structural measurement only, without measuring currents or carrier densities, which are prone to errors from Faradaic currents and migration of oxygen vacancies \cite{Jeong2013}. This result suggests that the carrier densities as large as several 10$^{13}$ cm$^{-2}$ observed in SrTiO$_3$ electric double-layer transistors \cite{Lee2011f} are consistent with electrostatic accumulation.

\subsection*{Methods}

Crystals of SrTiO$_3$ (channel) were immersed in an ionic liquid, and a gold electrode (gate) was used to apply a gate voltage. As shown in Figure \ref{fig:overview}(a), this system has 2 electrodes. The channel serves as the working electrode, and the gate serves as the counter electrode. The SrTiO$_3$ and gold electrodes were each 1 cm$^2$.

SrTiO$_3$ crystals (Shinkosha, STEP quality) were sonicated in acetone, then rinsed in isopropanol and de-ionized water, followed by a 2 min dip in 1:6 buffered oxide etch. As shown in the atomic force micrograph in Figure \ref{fig:overview}(c), this procedure yielded an atomically flat surface. Contact regions were defined by photolithography (S1813 photoresist, CD-30 developer). The contacts were argon ion milled (900 V accelerating potential, 0.5 mA/cm$^2$ beam current) for 60 seconds, and covered with 2 nm titanium and 30 nm gold (e-beam evaporation, 2 \AA/s). The surfaces were then cleaned with acetone, isopropanol, de-ionized water, and 15 seconds in an oxygen plasma asher (March PX250, remote plasma, 300 W, 300 mbar O$_2$).

The ionic liquid used in all experiments was 1-butyl-1-methylpyrrolidinium tris(pentafluoroethyl) trifluorophosphate (BMPY-FAP, 99 \% purity, EMD-Millipore). A tri-block co-polymer (PS-PMMA-PS, 86 kDa) was added at a weight concentration of 1:25 PS-PMMA-PS:BMPY-FAP in order to stabilize a thin ($\sim$ 1 $\mu$m) layer of ionic liquid on the SrTiO$_3$ surface \cite{Cho2008a}. As we show in the Supporting Information, the co-polymer has little effect on the interface. All measurements were performed in 10$^{-5}$ Torr vacuum at room temperature.

As shown in Figure \ref{fig:overview}(b), no Faradaic peaks were observed in cyclic voltammetry measurements, indicating the high purity of the ionic liquid. The SrTiO$_3$ became conductive at positive applied voltage, an observation which is consistent with previous experiments \cite{Lee2011f}, and which shows that the small amount of polymer in the ionic liquid does not significantly change the double layer properties.

Reflectivity was measured on beamlines 7-2 and 10-2 at the Stanford Synchrotron Radiation Lightsource using a Pilatus 100k area detector. The x-ray energy was 15.5 keV. The sample was translated 500 $\mu$m/min laterally to minimize beam damage (see Supporting Information for a discussion of beam damage). The samples were slightly miscut, allowing reflectivity to be collected as far as the substrate SrTiO$_3$ (001) peak.

\subsection*{Acknowledgments}

T. P. was supported by the Department of Defense (DoD) through the National Defense Science and Engineering Graduate Fellowship (NDSEG) Program and a Stanford Graduate Fellowship (SGF). Part of this work was performed at the Stanford Nano Shared Facilities (SNSF), supported by the National Science Foundation under award ECCS-1542152. This work is supported by the Department of Energy, Laboratory Directed Research and Development. X-ray reflectivity measurements were performed at Stanford Synchrotron Radiation Lightsource, SLAC National Accelerator Laboratory, which is supported by the U.S. Department of Energy, Office of Science, Office of Basic Energy Sciences under Contract No. DE-AC02-76SF00515.

\subsection*{}
\textbf{Supporting Information Available:} Reflectivity over a larger range of scattering vectors than in the main text, a discussion of beam damage, fits using other models than the one in the main text, and a table of best fit parameters. This material is available free of charge \emph{via} the Internet at http://pubs.acs.org.

\clearpage


\providecommand*\mcitethebibliography{\thebibliography}
\csname @ifundefined\endcsname{endmcitethebibliography}
  {\let\endmcitethebibliography\endthebibliography}{}

\end{document}